\documentclass[aip,rsi,reprint]{revtex4-1} 
\usepackage{graphicx}
\usepackage{subfigure}
\usepackage{epsfig}
\usepackage{color}

\begin{document}


\title{Setup for meV-resolution inelastic X-ray scattering measurements and X-ray diffraction at the Matter in Extreme Conditions Endstation at the Linac Coherent Light Source} 
\thanks{Paper published as part of the Proceedings of the 22nd Topical Conference on High-Temperature Plasma Diagnostics, San Diego, California, April, 2018.\\}

\author{E.~E.~McBride}
\email{emcbride@slac.stanford.edu}
\affiliation{SLAC National Accelerator Laboratory, 2575 Sand Hill Road, Menlo Park, California 94025, USA}
\affiliation{European XFEL GmbH, Holzkoppel 4, D-22869 Schenefeld, Germany}

\author{T.~G.~White}
\affiliation{University of Nevada at Reno, Reno, NV 89506 USA}

\author{A.~Descamps}
\affiliation{SLAC National Accelerator Laboratory, 2575 Sand Hill Road, Menlo Park, California 94025, USA}
\affiliation{Department of Aeronautics and Astronautics, Stanford University, Stanford, California 94305, USA}

\author{L.~B.~Fletcher}
\affiliation{SLAC National Accelerator Laboratory, 2575 Sand Hill Road, Menlo Park, California 94025, USA}

\author{K.~Appel}
\affiliation{European XFEL GmbH, Holzkoppel 4, D-22869 Schenefeld, Germany}

\author{F.~Condamine}
\affiliation{Sorbonne Universités, UPMC, LULI, UMR 7605, case 128, 4 place Jussieu 75252 Paris Cedex 05, France}
\affiliation{LULI, Ecole Polytechnique, CEA-CNRS-UPS, 91228 Palaiseau, France}

\author{C.~B.~Curry}
\affiliation{SLAC National Accelerator Laboratory, 2575 Sand Hill Road, Menlo Park, California 94025, USA}
\affiliation{Department of Electrical and Computer Engineering, University of Alberta, Edmonton, AB T6G 1H9, Canada}

\author{F.~Dallari}
\affiliation{Dipartimento di Fisica, Universit\`{a} di Trento, via Sommarive 14, 38123 Povo (TN), Italy}

\author{S.~Funk}
\affiliation{Friedrich-Alexander-Universit\"at Erlangen-N\"urnberg, Erlangen Centre for Astroparticle Physics, Erwin-Rommel-Str. 1, D-91058 Erlangen, Germany}

\author{E.~Galtier}
\affiliation{SLAC National Accelerator Laboratory, 2575 Sand Hill Road, Menlo Park, California 94025, USA}

\author{M.~Gauthier}
\affiliation{SLAC National Accelerator Laboratory, 2575 Sand Hill Road, Menlo Park, California 94025, USA}

\author{S.~Goede}
\affiliation{European XFEL GmbH, Holzkoppel 4, D-22869 Schenefeld, Germany}

\author{J.~B.~Kim}
\affiliation{SLAC National Accelerator Laboratory, 2575 Sand Hill Road, Menlo Park, California 94025, USA}

\author{H.~J.~Lee}
\affiliation{SLAC National Accelerator Laboratory, 2575 Sand Hill Road, Menlo Park, California 94025, USA}

\author{B.~K.~Ofori-Okai}
\affiliation{SLAC National Accelerator Laboratory, 2575 Sand Hill Road, Menlo Park, California 94025, USA}
\affiliation{Department of Chemistry, Massachusetts Institute of Technology, 77 Massachusetts Avenue, Cambridge, MA 02139, USA}

\author{M.~Oliver}
\affiliation{Department of Physics, Clarendon Laboratory, University of Oxford, Parks Road, Oxford OX1 3PU, UK}

\author{A.~Rigby}
\affiliation{Department of Physics, Clarendon Laboratory, University of Oxford, Parks Road, Oxford OX1 3PU, UK}

\author{C.~Schoenwaelder}
\affiliation{SLAC National Accelerator Laboratory, 2575 Sand Hill Road, Menlo Park, California 94025, USA}
\affiliation{Friedrich-Alexander-Universit\"at Erlangen-N\"urnberg, Erlangen Centre for Astroparticle Physics, Erwin-Rommel-Str. 1, D-91058 Erlangen, Germany}

\author{P.~Sun}
\affiliation{SLAC National Accelerator Laboratory, 2575 Sand Hill Road, Menlo Park, California 94025, USA}

\author{Th.~Tschentscher}
\affiliation{European XFEL GmbH, Holzkoppel 4, D-22869 Schenefeld, Germany}

\author{B.~B.~L.~Witte}
\affiliation{SLAC National Accelerator Laboratory, 2575 Sand Hill Road, Menlo Park, California 94025, USA}
\affiliation{Universit\"at Rostock, Institut f\"ur Physik, D-18051 Rostock, Germany}

\author{U.~Zastrau}
\affiliation{European XFEL GmbH, Holzkoppel 4, D-22869 Schenefeld, Germany}

\author{G.~Gregori}
\affiliation{Department of Physics, Clarendon Laboratory, University of Oxford, Parks Road, Oxford OX1 3PU, UK}

\author{B.~Nagler}
\affiliation{SLAC National Accelerator Laboratory, 2575 Sand Hill Road, Menlo Park, California 94025, USA}

\author{J.~Hastings}
\affiliation{SLAC National Accelerator Laboratory, 2575 Sand Hill Road, Menlo Park, California 94025, USA}

\author{S.~H.~Glenzer}
\affiliation{SLAC National Accelerator Laboratory, 2575 Sand Hill Road, Menlo Park, California 94025, USA}

\author{G.~Monaco}
\affiliation{Dipartimento di Fisica, Universit\`{a} di Trento, via Sommarive 14, 38123 Povo (TN), Italy}

\date{\today}

\begin{abstract}
We describe a setup for performing inelastic X-ray scattering measurements at the Matter in Extreme Conditions (MEC) endstation of the Linac Coherent Light Source (LCLS). This technique is capable of performing high-, meV-resolution measurements of dynamic ion features in both crystalline and non-crystalline materials. A four-bounce silicon (533) monochromator was used in conjunction with three silicon (533) diced crystal analyzers to provide an energy resolution of $\sim$50 meV over a range of $\sim$500 meV in single shot measurements.  In addition to the instrument resolution function, we demonstrate the measurement of longitudinal acoustic phonon modes in polycrystalline diamond. Furthermore, this setup may be combined with the high intensity laser drivers available at MEC to create warm dense matter, and subsequently measure ion acoustic modes. 
\end{abstract}

\maketitle
  
\section{Introduction}

Warm dense matter, an exotic state of matter found in fusion processes\cite{Glenzer2010} and in the cores of large planets and stellar accretion disks,\cite{Guillot1999} is too hot, and hence too highly ionized, to be described by condensed matter theories, and is too strongly coupled and correlated for classical plasma physics to provide an accurate description. A direct characterization of such a state is therefore vital. From an experimental point of view, this is challenging as the high densities of free electrons make warm dense matter opaque to visible light and, therefore, typical optical spectroscopic techniques are not possible. X-ray scattering diagnostics, such as X-ray diffraction or inelastic X-ray scattering, however, have shown great potential in diagnosing this extreme and complex state.\cite{Glenzer2003, Fletcher2015} 

Inelastic X-ray scattering refers to a broad range of measurement techniques involving energy and momentum transfer that allows one to measure the dynamical structure factor, S(Q, $\omega$), of a material. Techniques and instrumentation developed at 3rd generation synchrotron lightsources have allowed one to measure dynamic ion features in crystalline and non-crystalline matter, such as phonon modes in solids, and collective dynamics in disordered materials.\cite{Burkel2000, Sette1998} Such modes are analogous to collective excitations in warm dense matter. The measurement of these ion acoustic modes will allow the measurement of fundamental properties of the exotic warm dense matter state, such as sound velocity, viscosity, thermal conductivity.\cite{Rueter2014, Witte2017, Gregori2009, Mabey2017}

Here, we present an experimental setup for inelastic X-ray scattering based on the scheme described in Ref.\cite{Huotari2005} and fully adapted for measurements at free electron laser endstations. Specifically, the setup discussed here allows for measurements over a $\sim$500 meV exchanged energy range in single shot acquisition at an incident energy of $\sim$7.5 keV and with an energy resolution of $\sim$50 meV. It is thus capable of resolving inelastic phenomena in the tens of meV energy transfer range, of interest in studies of warm dense matter. The experimental setup was fielded at the Matter in Extreme Conditions (MEC) endstation of the hard X-ray free electron laser, the Linac Coherent Light Source (LCLS). The high-intensity long- and short-pulse lasers available at this endstation may be readily combined with the X-ray instrumentation, described below, to generate warm dense matter, and simultaneously measure the dynamic structure factor. We focus on a description of the X-ray scattering spectrometer, and demonstrate its utility by measuring phonon modes in polycrystalline diamond.

\begin{figure}[!ht]
\centering
\includegraphics[width=\columnwidth]{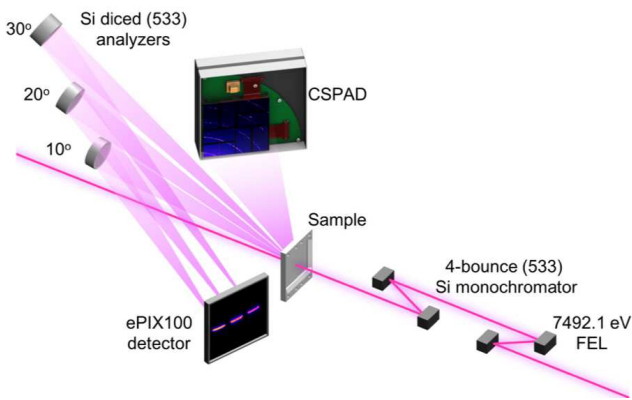}
\caption{\label{fig:setup} Experimental setup for performing meV-resolution inelastic X-ray scattering at the MEC endstation at LCLS.}
\end{figure}

\begin{figure}[!hb]
\centering
\includegraphics[width=0.8\columnwidth]{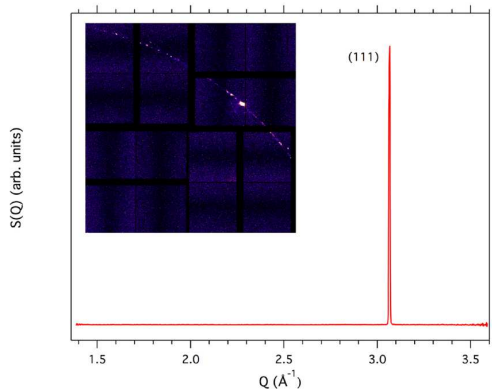}
\caption{\label{fig:diff} An integrated 1D diffraction pattern from polycrystalline diamond collected in a single X-ray pulse using the seeded beam of the LCLS, without monochromatization. One may see the sharp Bragg peak of the (111) reflection. Inset: 2D diffraction pattern.}
\end{figure}

\section{Experimental Setup}

Experiments were conducted at the MEC endstation\cite{Nagler2015} at the LCLS. A schematic of the setup used may be seen in Fig.~\ref{fig:setup}. The incident X-ray beam provided by LCLS had a temporal pulse length of $\sim$80\,fs and was focused to a spot size of 5\,$\mu$m on the target using a stack of beryllium compound refractive lenses. The LCLS was operated in the seeded mode, resulting in an incident X-ray beam with a bandwidth of $\Delta$E $\sim$ 0.5\,eV at an X-ray energy of 7492.1 eV. The X-rays were then monochromatized with a four-bounce channel-cut (533) silicon monochromator, resulting in a subsequent X-ray beam bandwidth of 32\,meV i.e. $\frac{\Delta E}{E}\sim4.3\times10^{-6}$. Following monochromatization, the number of photons per pulse was reduced from $\sim$10$^{11}$ in the seeded beam, to $\sim$10$^{10}$ photons on target. The corresponding intensity in the focus, on target, of the 5 $\mu$m, 80\,fs X-ray beam is then 7.6$\times$10$^{14}$ W/cm$^2$ for a single pulse.

\begin{figure}[!ht]
\centering
\includegraphics[width=0.7\columnwidth]{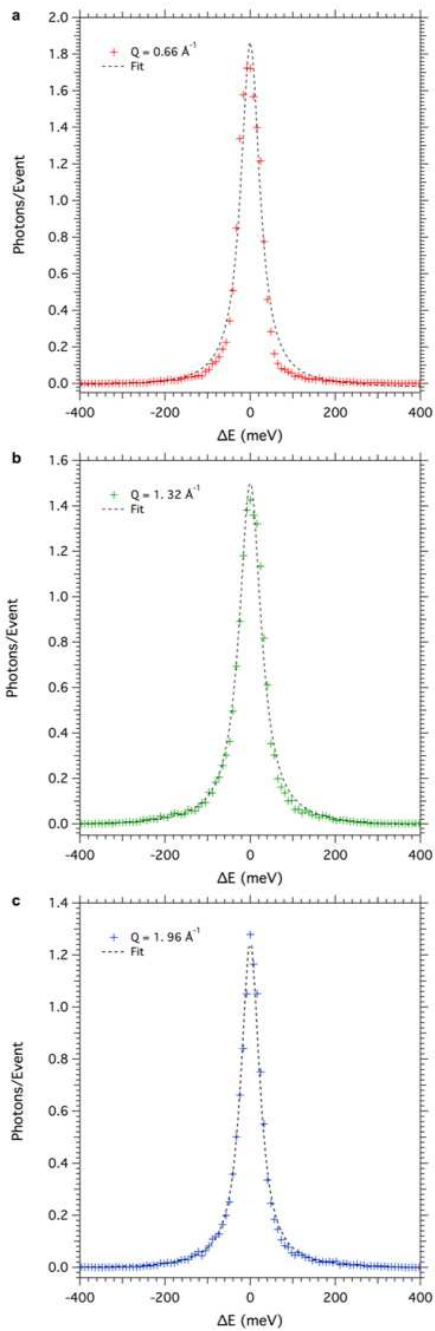}
\caption{\label{fig:resolution} Quasielastic scattering from 25 $\mu$m thick PMMA used to measure the resolution of each analyzer. Raw data are shown by the red, green and blue markers. Lorentzian fits to the data are shown by the dashed line in each figure. The energy resolution was measured to be {\bf a} $\Delta$E = 53\,meV at  Q = 0.66\,\AA$^{-1}$, {\bf b} $\Delta$E = 61\,meV at Q = 1.32\,\AA$^{-1}$, and {\bf c} $\Delta$E = 51\,meV at Q = 1.96\,\AA$^{-1}$}
\end{figure}

\begin{figure*}[!ht]
\centering
\includegraphics[width=\textwidth]{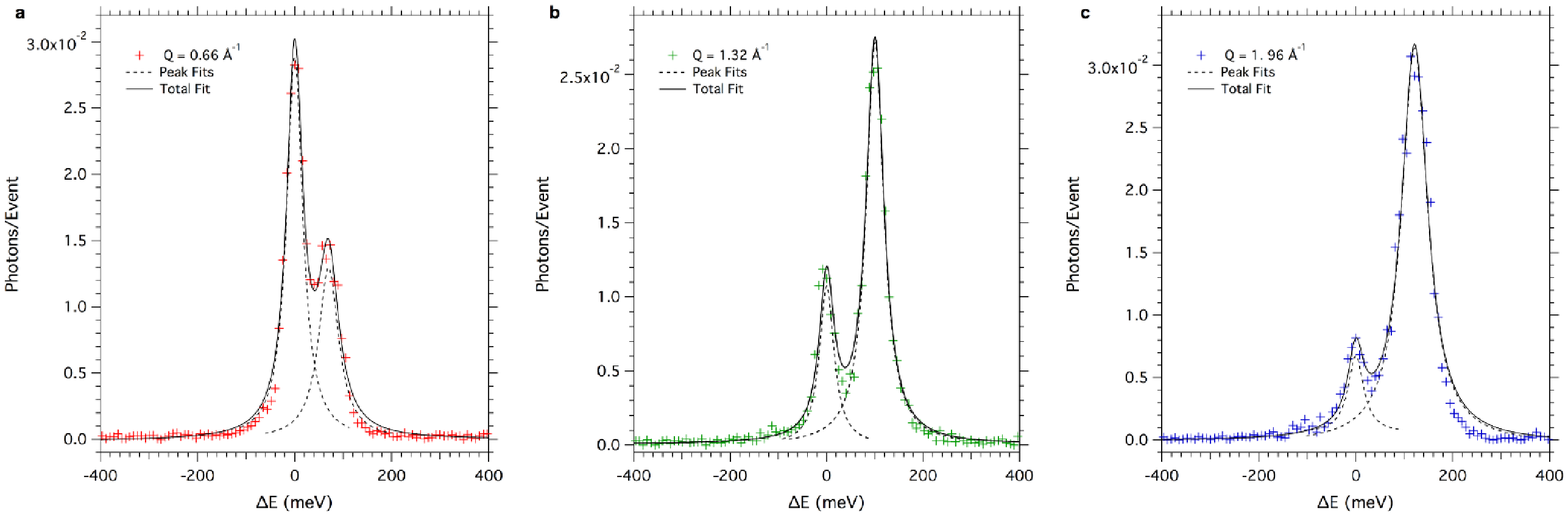}
\caption{\label{fig:diamond} Inelastic X-ray scattering spectra from polycrystalline diamond at \textbf{a} Q = 0.66\,\AA$^{-1}$, \textbf{b} Q = 1.32\,\AA$^{-1}$, and \textbf{c} Q = 1.96\,\AA$^{-1}$ are shown by the crosses. Dashed lines in each figure are Lorentzian peak fits to the quasielastic and inelastic scattering data, and the solid black lines are the total fit to the data.}
\end{figure*}

The monochromatized X-rays were then scattered from the sample onto the crystal analyzers, and subsequently onto the detector. Three, 100 mm diameter, diced silicon (533) crystal analyzers\cite{Verbeni2005} with a bending radius R = 1 m were used in a Johann geometry. The size of the diced cubes were 1.5 mm $\times$ 1.5 mm in order to guarantee that a sufficiently large spectrum could be measured in single shot. The analyzers were operated at a Bragg angle of  $\theta_B$ = 87.5$^\circ$ -- the same Bragg angle as the four-bounce channel-cut monochromator. In this configuration, the analyzer, the sample, and the detector are placed on the so-called Rowland circle, a circle with a radius which equals half the radius of curvature of the analyzer crystals. The analyzer crystals were positioned vertically at angles of 10$^\circ$, 20$^\circ$ and 30$^\circ$, corresponding to an exchanged scattering vector Q of 0.66\,\AA$^{-1}$, 1.32\,\AA$^{-1}$, and 1.96\,\AA$^{-1}$, respectively. The X-rays diffracted by the analyzers were detected on an ePIX100 detector with a pixel size of 50 $\mu$m $\times$ 50 $\mu$m and an active area of 38.4 mm $\times$ 35.2 mm.\cite{Blaj2015} Due to the low efficiency of the inelastic scattering process, the LCLS was operated at a repetition rate of 120 Hz, allowing for rapid accumulation of data. Data were subsequently processed using a single photon counting code developed by Condamine \textit{et al.}\cite{Condamine}

In addition to inelastic scattering measurements, a Cornell Stanford Pixellated Area Detector (CSPAD)\cite{Blaj2015} was used to collect X-ray diffraction patterns, allowing the measurement of the structure factor, S(Q). The detector was placed at a distance of 131.64 mm from the sample, covering a 2$\theta$ range of 20.7$^\circ$ to 56.6$^\circ$, corresponding to a Q-range of 1.36 \AA$^{-1}$ to 3.60 \AA$^{-1}$. The incident X-ray flux following monochromatization of the seeded beam is too low to collect single-exposure diffraction patterns. Consequently, the monochromator crystals were mounted on a motorized holder, allowing them to be moved reproducibly into and out of the X-ray beam, allowing one to switch between an inelastic scattering configuration, to a diffraction configuration. Indeed, the choice of a four-bounce channel-cut monochromator was motivated by the possibility of choosing the incident beam intensity (and resolution) with a fast moving assembly, leaving the beam position unaltered. This is crucial for future experiments on matter in extreme conditions where one wants to measure both diffraction and inelastic X-ray scattering from the same sample spot, overlapped with the laser driver. An example of an azimuthally integrated X-ray diffraction pattern from 40 $\mu$m thick polycrystalline diamond, the purpose of which is described in Section \ref{sec:diamond}, is shown in Fig. \ref{fig:diff}. Here, we clearly observe the (111) reflection of diamond at Q = 3.06 \AA$^{-1}$, corresponding to a lattice parameter of 3.55 \AA. 

\section{Instrument resolution}

The energy resolution of the instrument has many different contributions; it is a convolution of contributions related to the Darwin width of the monochromator and the analyzer crystal, to the beam size on sample, and the detector pixel size\cite{Huotari2005}. It was measured for each analyzer using the quasielastic scattering (zero energy transfer) from a 25\,$\mu$m thick piece of PMMA (Goodfellows Ltd.). A total of 1776 events were accumulated. Figure \ref{fig:resolution} shows the instrument function measured from all three analyzers, normalized to the number of events. The resolution of each analyzer, $\Delta$E, was determined following a Lorentzian fit to the data, also shown in Fig. \ref{fig:resolution}, and were measured to be $\Delta$E = 53\,meV, 61\,meV, and 51\,meV at Q = 0.66\,\AA$^{-1}$, Q = 1.32\,\AA$^{-1}$, and Q = 1.96\,\AA$^{-1}$, respectively. These numbers should be compared with an estimated value of 46 meV. The differences between the expected value and the measured ones might be due, at least in part, to some stress present in the diced crystals, and related to their large cube size when compared with more standard ones.\cite{Verbeni2005}

\section{Example: Measurement of Acoustic Phonons in Polycrystalline Diamond}\label{sec:diamond}

As a demonstration, we used the instrumentation described above to measure phonon modes in polycrystalline diamond. X-rays were incident onto a 3 mm $\times$ 3 mm square of polycrystalline diamond (Applied Diamond Inc.), 40 $\mu$m thick in the X-ray direction. Here, a total of 14124 events were collected. The crosses in Figs. \ref{fig:diamond} a-c show the integrated raw spectra for each analyzer, normalized to the number of events. Dashed lines shown in each figure are Lorentzian peak fits to the quasielastic and the inelastic X-ray scattering peaks. The solid black lines are total fits to the data. The momentum, Q, and energy, E, of the inelastic peak is listed in Table \ref{Table1}. Also listed is the frequency, $\nu$, of the phonon mode measured. 

\begin{table}
\begin{center}
\begin{tabular}{ c | c | c }		
 \hline	
 \hline	
Q \AA$^{-1}$ & E (meV) &  $\nu$ (THz) \\
 \hline	
0.66 & 74 & 17.9 \\
1.32 & 108 & 26.1 \\
1.96 & 126 & 30.5 \\
 \hline	
 \hline	
\end{tabular}
\caption{Measured phonon energies and corresponding frequencies as determined from the inelastic scattering data shown in Fig. 4.}\label{Table1}
\end{center}
\end{table}

\begin{figure}
\centering
\includegraphics[width=\columnwidth]{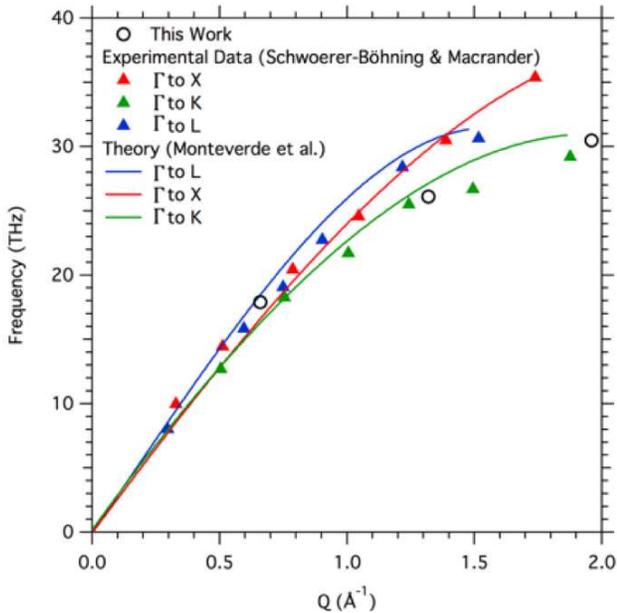}
\caption{\label{fig:Dispersion} Acoustic phonon frequencies measured in polycrystalline diamond. Open circles are phonon frequencies measured in this study. Triangles show inelastic X-ray scattering data from Schwoerer-B\"ohning \& Macrander. \cite{Schwoerer-Boehning1998} Solid lines are calculated phonon dispersion curves for longitudinal acoustic phonons in single crystal diamond in the $\Gamma$ to K, L and X directions, as calculated by Monteverde \textit{et al.} \cite{Monteverde2015}}
\end{figure}

Figure \ref{fig:Dispersion} shows a comparison of the measured phonon frequencies from polycrystalline diamond with longitudinal acoustic phonon modes measured from single crystal diamond using inelastic X-ray scattering at a synchrotron light source.\cite{Schwoerer-Boehning1998} The data are further compared with molecular dynamics simulations by Monteverde \textit{et al.},\cite{Monteverde2015}. The phonon modes we measure in the polycrystalline diamond sample are shown to be consistent with longitudinal acoustic modes.

\section{Summary}

In this paper we have presented an experimental platform to perform meV-resolution inelastic scattering experiments at the MEC endstation of the LCLS. The instrument resolution was measured using quasielastic scattering from PMMA, and was determined to be $\sim$50 meV over an exchanged energy range of $\sim$500 meV for single shot measurements. In addition, we demonstrate the measurement of longitudinal acoustic modes in polycrystalline diamond. Coupling this experimental configuration with high intensity long- and short-pulse lasers available at MEC, LCLS, USA, or the soon to be commissioned High Energy Density (HED) instrument at European XFEL, Germany,\cite{Tschentscher2017} will provide an avenue to create and diagnose ion acoustic modes in warm dense matter.

\begin{acknowledgments}
E. E. McBride acknowledges funding from the Volkswagen Foundation. Support from AWE plc., the Engineering and Physical Sciences Research Council (grant numbers EP/M022331/1 and EP/N014472/1) and the Science and Technology Facilities Council of the United Kingdom is also acknowledged. The ESRF is gratefully acknowledged for the provision of the four-bounce monochromator. Use of the Linac Coherent Light Source (LCLS), SLAC National Accelerator Laboratory, is supported by the U.S. Department of Energy, Office of Science, Office of Basic Energy Sciences under Contract No. DE-AC02-76SF00515. The MEC instrument is supported by the U.S. Department of Energy, Office of Science, Office of Fusion Energy Sciences under contract No. SF00515. SLAC High Energy Density Science Division acknowledges support from FES FWP100182.
\end{acknowledgments}


\end{document}